\documentclass[12pt]{article}

\usepackage{epsfig}
\usepackage{cite}
\usepackage{amsmath, amssymb, amsfonts}
\usepackage{color}
\usepackage{latexsym}
\usepackage{graphicx}

\def\be{\begin{equation}}
\def\ee{\end{equation}}
\def\ba{\begin{eqnarray}}
\def\ea{\end{eqnarray}}

\def\bdm{\begin{displaymath}}
\def\edm{\end{displaymath}}
\def\la{~\mbox{\raisebox{-.6ex}{$\stackrel{<}{\sim}$}}~}
\def\ga{~\mbox{\raisebox{-.6ex}{$\stackrel{>}{\sim}$}}~}
\def\bq{\begin{quote}}
\def\eq{\end{quote}}

 at 10truept

\newcommand{\p}{\partial}

\newcommand{\eps}{\epsilon}

\newcommand{\Mpl}{m_{\mathrm{Pl}}}

\newcommand{\bea}{\begin{eqnarray}}
\newcommand{\eea}{\end{eqnarray}}

\newcommand{\half}{\frac{1}{2}}
\newcommand{\bi}{\begin{itemize}}
\newcommand{\ei}{\end{itemize}}

\newcommand{\beq}{\begin{equation}}
\newcommand{\eeq}{\end{equation}}
\newcommand{\beqa}{\begin{eqnarray}}
\newcommand{\eeqa}{\end{eqnarray}}
\newcommand{\mpl}{\Mpl}

\def\la{~\mbox{\raisebox{-.6ex}{$\stackrel{<}{\sim}$}}~}
\def\ga{~\mbox{\raisebox{-.6ex}{$\stackrel{>}{\sim}$}}~}

\newcommand{\Z}{\mathbb{Z}}

\def\ltap{\ \raise.3ex\hbox{$<$\kern-.75em\lower1ex\hbox{$\sim$}}\ }
\def\gtap{\ \raise.3ex\hbox{$>$\kern-.75em\lower1ex\hbox{$\sim$}}\ }
\def\gl{\ \raise.5ex\hbox{$>$}\kern-.8em\lower.5ex\hbox{$<$}\ }
\def\roughly#1{\raise.3ex\hbox{$#1$\kern-.75em\lower1ex\hbox{$\sim$}}}

\begin{document}

\thispagestyle{empty}
\begin{flushright}
September 2018 \\
BRX-TH-6637\\
CERN-TH-2018-202
\end{flushright}
\vspace*{.75cm}
\begin{center}

{\Large \bf Strongly Coupled Quintessence}

\vspace*{1cm} {\large Guido D'Amico$^{a, }$\footnote{\tt
damico.guido@gmail.com}, Nemanja Kaloper$^{b, }$\footnote{\tt
kaloper@physics.ucdavis.edu} and Albion Lawrence$^{c, }$\footnote{\tt albion@brandeis.edu}
}\\
\vspace{.3cm} {\em $^a$Theoretical Physics Department,
CERN, Geneva, Switzerland}\\
\vspace{.3cm}
{\em $^b$Department of Physics, University of
California, Davis, CA 95616, USA}\\
\vspace{.3cm} {\em $^c$Martin Fisher School of Physics, Brandeis University, Waltham, MA 02453, USA}\\

\vspace{1cm} ABSTRACT
\end{center}
We present a family of consistent quantum field theories of monodromy quintessence in strong coupling, 
which can serve as benchmarks in modeling dark energy different from cosmological constant. These theories
have discrete gauge symmetries which can protect them from quantum field theory and quantum gravity corrections, both perturbative and nonperturbative. The strong coupling effects, at scales $\ga {\rm mm}^{-1}$, flatten the potential and activate operators with higher powers of derivatives. The predicted  equation of state is close to, but not exactly equal to $-1$, thus being within reach of the 
(near!) future programs to explore the nature of dark energy. 
\vfill \setcounter{page}{0} \setcounter{footnote}{0}
\newpage

Roughly three quarters of the invisible world is dark energy, whose dynamics is not understood. It may be a cosmological constant, but explaining how it would be as small as needed is a well-known challenge: one needs some reason to ignore or almost completely cancel the large quantum vacuum energy contributions 
\cite{zeldovich,wilczek,wein}. The  alternatives that treat dark energy as a dynamical field, a.k.a quintessence, are even more challenging: one needs both the magnitude {\it and} the slope of the potential to be exquisitely small compared to the Planck scale or any fundamental scales of the Standard Model.

That being said, quintessence is a simple concept, and future observations of the expansion history of the universe will probe a large and interesting range of parameters. It is important to better understand whether a microscopic theory of quintessence can be made consistent and, to any degree possible, natural (in the sense of Wilson and 't Hooft). In this letter we discuss these issues and provide a class of models which are natural and appear to consistently couple to quantum gravity. Regardless of whether quintessence is realized in nature, a discussion of these issues and their resolution turns out to be interesting in its own right.

For a canonically normalized quintessence field with scalar potential $V(\phi)$, such that $V(0) = 0$, we must satisfy two constraints.  First, the vacuum energy at the present epoch must be consistent with the present Hubble constant, that is,
$V \sim (2\times 10^{-30} m_{pl})^4$, where $m_{pl}$ is the reduced Planck mass.  Secondly, the equation of state parameter $w$, defined by $p = w \rho$, is related to the slope of the potential by:
\be
	w = \frac{\eps/3 - 1}{\eps/3 + 1}\, , \qquad \qquad
	\eps = \frac{m_{pl}^2}{2}\left(\frac{V'(\phi)}{V(\phi)}\right)^2 \, .
\ee
Observations indicate\footnote{We ignore the more exotic situation $w<-1$ which can nevertheless be realized without pathologies \cite{wminone}.} that $-1 \leq w \leq-0.95$, or $0 \leq\eps\leq 0.075$. During the observable epoch, the quintessence field should traverse a distance $\delta\phi \sim {\dot\phi}H_0^{-1} \sim \sqrt{\eps} m_{pl}$, where we have used the slow roll equations.

Writing down models that satisfy these constraints requires some care when we take quantum gravity into account. The simplest potentials, including ones which are technically natural from a QFT point of view, require that $\phi$ is at a distance $\Delta \phi > \mpl$ from the minimum \cite{frieman,barbieri,hill,yanagida,KSQ,nilles,trivedi}. When coupled to quantum gravity, fairly basic arguments render such field ranges inaccessible to a single, simple effective field theory (EFT).  This is a slightly different problem than the one large-field inflation faces \cite{Linde:1984ir}: in single field models the inflaton must traverse super-Planckian distances {\it during} inflation, and physics over those field ranges would be imprinted in the observable sky. In contrast, as we will see, the quintessence scalar need only traverse sub- to near-Planckian distances {\it during} the observable epoch. Whether it really needs to change by a Planckian scale is
a question of model-building and eschatology. 

More seriously, as with large-field inflation, the potential slopes required by quintessence must be small\footnote{One would in principle have to worry about direct couplings of such light fields to matter because they could mediate long range forces \cite{KSQ}.}. One can try to protect this with a global shift symmetry, but such symmetries are violated by quantum gravity effects (see \cite{weak} for a modern viewpoint). Planck-suppressed operators with order ${\cal O}(1)$ coefficients then spoil the required properties of the theory.  An infinite number of fine tunings are required to maintain the small slope and intercept of the quintessence potential, absent a mechanism which suppresses even quantum gravity effects.

Finally, in addition to the small slope of the potential, dynamical dark energy models should in principle explain the near-cancellation of the vacuum energy, such that the present dark energy is dominated by the excess potential energy of the quintessence field. For effective field theories extending to the minimum of the potential, one must solve the cosmological constant problem. We will not attempt to address this notorious problem, but will merely assume the vacuum energy is somehow cancelled at the minimum of the quintessence potential\footnote{An amusing possibility is that the small vacuum energy is explained by the existence of a corner of the string landscape consistent with the existence of physicists who can argue about it.  A more amusing possibility is that the string landscape contains both quintessence and metastable de Sitter space.}.  We must still find a sensible EFT with a sufficiently small slope over the required field range. This is difficult enough, and many commonly encountered quintessence proposals fail at this step. Let us review some simple examples before turning to our own proposal. 

Exponential potentials $V \sim \exp(\alpha \phi/\mpl)$ are popular, but to the extent they are meaningful, they are problematic as EFTs coupled to quantum gravity.  The constraints on quintessence require $\alpha < \sqrt{2}$, and 
the sub-Planckian bound on the field range which $\phi$ has traversed until now, having started to roll in the early universe, means that observations will probe at most a few low orders in a Taylor expansion in $\phi - \phi_*$ around some point $\phi_*$. All of the coefficients must be small, and a principle is required to ensure this (supersymmetry, broken above a TeV, will be of limited help).  If there is some mechanism enforcing the exponential form, the future evolution would be sensitive to it over super-Planckian distances. If the exponential potential were to dominate the evolution of the universe forever, it would yield eternal quintessence which would have cosmological event horizons; it is not clear such spacetimes make sense in a theory of quantum gravity \cite{hks,willy}. On the other hand, absent some additional mechanism, we expect the 4d EFT to change over super-Planckian field ranges. This implies that an exponential potential is not obviously meaningful, essentially reducing any worry over eternal quintessence. A  change to the potential could well terminate the slow roll needed to yield cosmological horizons.

A simple quadratic potential $V = \half m^2 \phi^2$ is technically natural, having an approximate shift symmetry for small $m$ which protects the theory from large perturbative corrections due to fields including the graviton running in loops (see \cite{KS} for a review.) However, the criteria above require that the value of $\phi$ during the observable epoch is $\phi_{obs} \ga 3 \mpl$. As with the exponential potential, absent some protection mechanism, such a field is expected to be governed by an EFT at $\phi = \phi_{obs}$ that is quite different from the theory at $\phi = 0$. But if this is true, we cannot expect the quadratic potential to be meaningful on its own. In practice, the best one can say is that $V = M^4 + g\delta\phi^p \ldots$, where $M$ is a constant, $\delta\phi$ the deviation of the potential around some expansion point, and $p$ some power which is generically expected to be unity. We are still left searching for a good explanation for the small size of $M, g$. Note that quadratic potentials eventually always fall out of slow roll, and so avoid the conceptual problems of eternal, exponential inflation.

Periodic pseudoscalars, aka axions, are protected from perturbative corrections by the topology of field space, and in the dilute instanton gas approximation, the instantons can easily be kept small implying that nonperturbative corrections are also small \cite{frieman,barbieri,hill,yanagida}. However, for a cosine potential $V \sim \mu^4 \cos(\phi/f)$, we must have $f > 2 m_{pl}$, $\Delta\phi > 1.65 m_{pl}$ to get $w$ in the right range \cite{liddle,goobar} (with $\phi < \pi f_\phi$ in order to avoid the quintessence tachyonic instability near the maximum of the cosine \cite{nessy}). Super-Planckian axion decay constants, however, render the theory unstable to nonperturbative quantum gravity effects which spoil the simple cosine potential by allowing for arbitrary harmonics with coefficients of the same order as the fundamental cosine potential.

A solution is hinted at by the fact that the dilute instanton gas approximation does not always work (it appears to fail even for $SU(3)$ Yang-Mills \cite{Giusti:2007tu}). The generic alternative, first noted in large-N theories \cite{Witten:1980sp}, is a multivalued potential, termed ``axion monodromy'' in the string theory literature \cite{eva}. In such theories, the effective field range can be super-Planckian. The EFT of these models \cite{KS,KSQ,NDA}\ demonstrates that a combination of continuous and discrete gauge symmetries protects the shift symmetry from large quantum gravity effects, even when the effective scalar appears to have super-Planckian expectation values. In this work, we adapt the general framework of \cite{KS,NDA}\ to describe theories of quintessence.

We start with the description of the axion as a dual massive $4$-form\footnote{We eliminate terms with $\p^k F$ factors using the equations of motion.}: 
\ba 
{\cal L}^{\rm (full)}  &=&  - \frac{1}{48} F_{\mu\nu\lambda\sigma}^2 - \frac{m^2}{12} (A _{\mu\nu\lambda} - h_{\mu\nu\lambda})^2   - \sum_{n>2} \frac{a'_n}{M^{2n-4}}\tilde F^{n}  \nonumber\\
& & - \sum_{n > 1} \frac{a''_n }{M^{4n-4}} m^{2n} (A_{\mu\nu\lambda} -h_{\mu\nu\lambda})^{2n}   - \sum_{k\ge1, \, l\ge 1} \frac{a'''_{k , l}}{M^{4k+2l-4}} m^{2k}
(A_{\mu\nu\lambda}-h_{\mu\nu\lambda})^{2k} \tilde F^{l}  \, . 	
\label{corrections}
\ea
Here $A$ is the gauge field 3-form, $F=dA$, ${\tilde F} = {}^*F$, $b$ a two-form St\"uckelberg gauge field with field strength $h=db$, and $m$ plays the role of both the gauge field mass and the Stueckelberg mode coupling. 
By gauge symmetry and the Goldstone Boson Equivalence Theorem (GBET), any power of $A$ not covered by a derivative must be multiplied by the same power of $m$ \cite{KS,NDA}. Finally, $M$ is the cutoff of the low energy EFT. This theory has a compact $U(1)$ {\it gauge} symmetry for the $4$-form and a discrete phase space gauge symmetry for the dual scalar. As a result the EFT (\ref{corrections}) is a full description of the dynamics below the cutoff $M$, satisfying technical naturalness and protected from quantum gravity corrections even when $m \ll  M$. 
The dimensionless coefficients $a_n', a_n'', a_{k,l}'''$ are fixed by naturalness: up to combinatorial factors and the loop factors they are\footnote{Up to factors which are logarithmic in momenta.} ${\cal O}(1)$ unless they are prohibited by symmetries, in order to guarantee that the action (\ref{corrections}) is complete. 

Next, we dualize the longitudinal mode of $F$ to a compact scalar: $F \sim \epsilon (m \phi + Q), mA \sim \epsilon \partial \phi$ \cite{KS,NDA}.  Here $Q = N q$, where $q$ is the fundamental 4-form charge; $N \in \Z$; and $\phi \equiv \phi + f$ where $mf = q$ \cite{KS}.  Defining the effective quintessence field $\varphi = \phi +Q/m$, and using Na\"ive Dimensional Analysis (NDA) \cite{georgi} to provide proper numerical
normalizations of the  dimensionless coefficients in (\ref{corrections}), which leads to substitutions $\varphi \rightarrow 4\pi \varphi/M$, inclusion of the factorials in the coefficients of (\ref{corrections}) to 
reproduce the symmetry factors of the S-matrix elements derived from (\ref{corrections}), and normalizing of the action by $M^4/(4\pi)^2$ following \cite{georgi}, we obtain after straightforward manipulation
\be \label{kinflation}
{\cal L} =  K\Bigl(\varphi, X \Bigr) - V_{eff}(\varphi)  =  \frac{M^4}{16\pi^2} {\cal K}\Bigl(\frac{4\pi m \varphi}{M^2}, \frac{16\pi^2  X}{M^4} \Bigr) - \frac{M^4}{16\pi^2} {\cal V}_{eff}\Bigl(\frac{4\pi m \varphi}{M^2}\Bigr) \, ,
\ee
where $X \equiv - (\partial_\mu \varphi)^2$, and ${\cal K}, {\cal V}_{eff}$ stand for asymptotic series, well approximated by finitely many terms, whose coefficients in the expansion are ${\cal O}(1)$ when the appropriate combinatorial factors are included, unless they vanish because of symmetries. This theory has a weakly broken shift symmetry $\phi \to \phi + \eps$, which for $f < m_{pl}$ is protected from further breaking by pertubative and nonperturbative effects as well as quantum gravity effects such as wormholes or intermediate black hole states, due to the gauge symmetries of the theory.

Just as the similar action applied to inflation was an example of the k-inflation models of \cite{ArmendarizPicon:1999rj},
the action (\ref{kinflation}) with a mass $m \sim H_0 \sim  10^{-33} \, {\rm eV}$ and a strong coupling cutoff $M \sim \sqrt{H_0 \mpl} \sim 10^{-3} \, {\rm eV}$ is a generalization of the phenomenological theory of k-essence first proposed in \cite{kessence}. Clearly, some dynamics in the hidden sector is required to generate such small scales, and many examples have been produced over time \cite{frieman,barbieri,hill,yanagida,KSQ,trivedi}. From our point of view, it is more important that once such small scales are generated, they are automatically protected by symmetries of the theory from dangerous corrections. What remains is to check that the models (\ref{kinflation}) pass the experimental bounds on dark energy \cite{liddle,goobar} while evading any additional constraints from quantum field theoretic or quantum gravity corrections.

In (\ref{kinflation}), while $M$ denotes the UV cutoff above which we must include additional degrees of freedom, $M_s = M/\sqrt{4\pi}$ is a strong coupling scale controlling the expansion of ${\cal L}$ in powers of $X,m\varphi$.  This action thus has two regimes \cite{NDA}: a weakly coupled regime $m\varphi/M_s^2, X/M_s^4 \ll 1$, and strongly coupled regimes for which either $M_s^2 \ll m\varphi \ll M^2$ or $M_s^4 \ll X \ll M^4$ (or both!). Note, that the regime
where the derivative terms are large but the field {\it vev} is small reduces to the conventional k-essence \cite{kessence}.  
 
In the weakly coupled regime the kinetic term is effectively canonical, and the potential effectively quadratic:
\be\label{eq:weaklag}
	{\cal L} = \half (\p\varphi)^2 - \half m^2\varphi^2
\ee
As noted above, in this regime the constraints $-1 < w < -0.95$ demand $\varphi > 3 m_{pl}$ compared to the minimum of the potential.  For $f < m_{pl}$ this involves some number of windings of $\varphi$, or alternatively a number of quanta of $F$.  While monodromy protects this regime from direct quantum corrections such  as wormholes or black hole intermediate states, it has been argued that in any UV completion the backreaction of light fields, or the inevitable appearance of new light fields, will alter the effective theory as the effective scalar $\varphi$ traverses a Planck distance \cite{vafao,reece,palti,valenzuela}. We remain agnostic about these arguments. As we will see, there is a natural cutoff in the strong coupling region of the theory, which may be compatible with the above arguments.  But it is not clear whether they will be relevant for the moderate distances traversed in the case of quintessence, or whether they would apply to the weak coupling regime -- explicit string constructions on which these arguments are based are hard to construct, for reasons outlined in \cite{Kaloper:2014zba}. 

The next step is to study the different regimes of strong coupling.  We first focus on the case that $X \ll M_s^4$, $M_s^2 < m\varphi < M^2 = 4\pi M_s^2$, which is self-consistent \cite{NDA}.  In this regime, the potential can flatten considerably in the strong coupling regime, consistent with string \cite{eva}\ and field-theoretic \cite{Dubovsky:2011tu}\ constructions. Note that the field space in our effective theory does have a cutoff at $\varphi = M^2/m$, at which point UV degrees of freedom are expected to become important. However, as we will see, there is ample room for quintessence to occur (just as there is room for inflation to occur in \cite{NDA}).

If the observable epoch occurs in the strong coupling regime, we can have a situation in which the weak coupling regime at the origin of field space is governed by a larger mass than above, and so the desired vacuum energy is reached at a smaller value of $\varphi$, near which the potential flattens enough to satisfy the bounds on $\eps$/$w$.  Let us consider a particular example to make our point. Start with
\be
	{\cal L} = k\left(\frac{m\varphi}{M_s^2}\right)(\p\varphi)^2 - M_s^4{\cal V}_{eff}\Bigl(\frac{m \varphi}{M_s^2}\Bigr)
\ee
In this single field, two-derivative example, we can find a new scalar $\chi$ which is canonically normalized:
\be
	{\cal L} = (\p\chi)^2 - M_s^4{\overline {\cal V}}_{eff}\Bigl(\frac{m \chi}{M_s^2}\Bigr)
\ee
Suppose the potential is (motivated by the extreme flattening case discussed by \cite{Dubovsky:2011tu})

\be
	{\overline {\cal  V}}(x) = 1 - \frac{1}{1 + x^2} \, .
\ee
The value of $m$ will be fixed by the location in field space where the observable epoch lies, and the vacuum energy and  choice of $w$ there.  If we adjust the parameters of the model to push our observable universe closer to the unitarity bound $\varphi = M^2/m$, we increase the value of $m$ and reduce the distance in Planck units of $\varphi$ today from $\varphi = 0$.  At the boundary, we find $\varphi = 0.1 \, m_{pl}$. Note that monodromy (or some other mechanism) is still required to suppress operators of the form $\varphi^p/m_{pl}^{p-4}$: if these had ${\cal O}(1)$ coefficients in the full EFT defined about $\varphi = 0$, they would spoil the desired slow-roll properties needed for quintessence. At any rate, here we have a model for which the full evolution of the universe can be governed by a single effective field theory ranging over sub-Planckian field ranges, with a quintessence potential which eventually falls out of slow roll.

Retention of the higher derivative terms, giving a form of k-essence  \cite{kessence}\ can further help maintain the slow roll regime. For specificity, consider a simple example with $K = {\cal Z} X + \tilde {\cal Z} {X^2}{/M^4_s}$; we offer this not as a realistic model but as an indication of how higher derivatives might operate.  In strong coupling $X \gg M_s^4$ and $ V_{eff} \gg M_s^4$ (we return to the un-normalized potential $V_{eff}$ for the convenience of comparing to observations), and so the leading order slow roll equations are
\be
3 \mpl^2 H^2 =  V_{eff} \, , ~~~~~~~~~~~~~  3H \dot \varphi \tilde {\cal Z} X = - 8\pi^2 M_s^2 m V_{eff}' \, .
\label{background}
\ee
When the potential is not too flat, if the derivative terms are
turned on they will remain in control for a few efolds at least, and may dominate over the quadratic derivative
terms all the way to the boundary of strong coupling, while driving cosmic acceleration. Indeed, manipulating\footnote{Using $3\mpl^2 H^2 = \rho$ and $\dot \rho = - 3H(1+w) \rho$.}
 (\ref{background}) we can derive
 \be 
 1+w \simeq \frac{8\pi^2}{9 \tilde {\cal Z}} \, \Big(\frac{V_{eff}'}{V_{eff}} \Big)^2 \, \frac{m^2 \mpl^2}{M_s^4} \, \frac{M_s^4}{X} \, , 
 \label{eqstderivs}
 \ee
 where the right hand side can be very small when\footnote{Note, that in the strong coupling regime,
 $X/M_s^4$ can be as large as $16\pi^2 \simeq 158$.} $X/M_s^4 \gg 1$ even when $V_{eff}' \la V_{eff}$ and
 $m \mpl \la M_s^2$, implying that $w \simeq -1$. As the field rolls the potential diminishes to $V_{eff} \simeq M^4_s$. If the weak coupling potential is not too shallow, with $m \mpl \sim M_s^2$,\footnote{At weak coupling this is the boundary of slow roll: at strong coupling, flattening and higher-derivative terms can maintain slow roll even at this boundary.} this violates the slow roll conditions and ends cosmic acceleration. Since this stage is short, the nonlinearities induced by higher derivatives at the large scales of the universe will not affect the background significantly \cite{guidosound} and the theory will remain consistent with observations.
However, higher derivatives will affect the growth of quintessence perturbations, resulting in a speed of sound smaller than unity. Notice from Eq. (\ref{eqstderivs}) that the larger the higher derivative operators, the equation of state 
parameter is closer to $w = -1$. On the other hand the speed of sound gets smaller. Hence in principle the perturbations could differentiate between strongly coupled quintessence and a cosmological constant.  
This is very interesting observationally, and constraints will be put in the near future~\cite{euclid}.

Note that in our discussion, the scales $m$ and $M$ are quite low. When we substitute the numerical scales of the dark energy and the Hubble parameter, $V_{eff} \simeq 10^{-12} \, {\rm eV}^4$ and $H_0 \simeq 10^{-33} \, {\rm eV}$, we find that the cutoff is 
\be
M \simeq \sqrt{4\pi} V_{eff}^{1/4} \simeq 3 \times 10^{-3} \, {\rm eV} \, , 
\ee
and so the quintessence mass is, when $m \sim M_s^2/\mpl \sim M^2/4\pi \mpl$,
\be
m \ga H_0 \simeq 10^{-33} \, {\rm eV} \, .
\ee
At the scales $M \ga {\rm mm}^{-1}$ there must be new physics in the dark sector that affects quintessence dynamics. It is intriguing to imagine that if this new physics is gravitationally coupled as we would expect, it could even generate corrections to gravity at sub-millimeter distances.

Recently it has been argued that (metastable) de Sitter vacua do not exist in string theory \cite{steinvafa}, and that string theory cosmologies with dynamical scalar fields only admit positive potentials while the fields are rolling, such that
\be
\frac {\mpl \, \partial_\varphi V}{V} > c \sim O(1) \, ,
\label{pot}
\ee
Much followup work has appeared since, with some examples given in  \cite{stuff}. This is a very strong constraint, and is not without criticism \cite{shamit}.  We have nothing to say about this condition, beyond noting that the phenomenological upper bound of $w = -0.95$ on the equation of state gives $c\sim \sqrt{2\eps}\sim 0.4$. Any further limits, or a measurement yielding $w$ much closer to $-1$, will put serious pressure on this proposal. For other recent works on quintessence vs ``the swampland" see e.g. \cite{Chiang:2018jdg,marsh,hitoshi,aklinde}.

To summarize, we have shown that the EFT of flux monodromy in strong coupling can naturally accommodate quintessence that easily meets the current observational limits on dark energy. The quintessence dynamics is stable
under quantum corrections, since the unbroken gauge symmetries of the dual massive $4$-form formulations serve as a protection mechanism. The resulting class of theories provides very useful benchmarks for future exploration of the nature of dark energy, giving a parameter space for the dark energy observables, in particular the equation of state $w$, consistent with quantum field theory and quantum gravity.

\section*{Acknowledgements}

N.K. thanks the CERN Theory Division for hospitality at the beginning of this work. 
N.K. is supported in part by DOE Grant DE-SC0009999. A.L. is supported in part by DOE grant DE-SC0009987.

\end{document}